\documentclass[fleqn,twoside]{article}
\usepackage{espcrc2}

\usepackage{graphicx}
\usepackage[figuresright]{rotating}

\newcommand{\AmS}{{\protect\the\textfont2
  A\kern-.1667em\lower.5ex\hbox{M}\kern-.125emS}}
\newcommand{\numu}[0]{\nu_\mu}
\newcommand{\nue}[0]{\nu_e}

\title{NuFact05 Working Group 2 Summary: Experimental Results in Neutrino 
       Scattering Physics}

\author{G.P. Zeller\address[COL]{Columbia University, Department of Physics,
        New York, NY 10027}%
        \thanks{gzeller@fnal.gov}}
 
\begin{document}

\begin{abstract}

This paper summarizes the experimental results in neutrino scattering physics 
as presented in Working Group 2. A summary of the theoretical results is 
presented in~\cite{sakuda-summary}.
\vspace{1pc}
\end{abstract}

\maketitle

\section{Introduction}

Current atmospheric and accelerator-based experiments study low energy
neutrino interactions to optimize their sensitivity to neutrino oscillations.
Maximum sensitivity requires precise knowledge of the cross sections for
charged current (CC) and neutral current (NC) neutrino-nucleus interactions
in the sub-to-few GeV range. The most relevant inputs include
quasi-elastic (QE) scattering cross sections and kinematics for accurate 
prediction of signal rates in these experiments. In addition, reliable
NC $\pi^0$ production cross sections are needed for estimation of the 
dominant backgrounds for $\numu \rightarrow \nue$ searches, CC $\pi^+$ 
production rates for estimation of the dominant 
backgrounds for $\numu$ disappearance searches, and antineutrino cross sections
for both signal and background predictions in CP violation searches. 
Because intense accelerator-based low energy neutrino beams have only 
been in operation for a few years, the information on these low 
energy neutrino cross sections comes primarily from bubble chamber, spark 
chamber, and emulsion experiments that ran decades ago. While 
extremely valuable for constraining present neutrino simulations, many of 
these experimental results suffered from low statistics and large 
neutrino flux uncertainties. As a result, many of these low energy cross 
sections are only known at the $10-40\%$ level compared to deep inelastic 
scattering cross sections which have been measured to a few-percent at 
higher energies.

Neutrino scattering physics has ventured into a new era. Recent data 
samples collected by both K2K and MiniBooNE have increased the 
statistics of this low energy data collection by almost an order of magnitude.
In doing so, both experiments have added important new data on nuclear 
targets ($^{16}O$, $^{12}C$) that are most relevant for use in neutrino
oscillation experiments. Higher energy data from the MINOS near 
detector~\cite{marchionni} that is just arriving will complement 
this picture. In addition, much anticipated plans for future measurements 
at the MINERvA experiment as well as plans to use beams of antineutrinos, 
stopped pion sources, beta beams, and liquid argon detectors are also on 
the horizon. These bring hope for rapid advancement in the field 
of neutrino scattering physics in the next decade or so. \\

The following sections briefly summarize the experimental results presented 
in this working group on these two fronts: recent measurements, and plans 
for future experiments.

\section{Status of Current Measurements}

Several exciting results made their debut at this workshop. 
The first is a search for CC coherent $\pi^+$ production using the fully 
active scintillator detector in the K2K near detector 
ensemble~\cite{k2k-coherent}. In the case of coherent single pion production, 
the neutrino scatters from the entire nucleus transferring negligble energy 
to the target. Hence, such interactions have the distinct signature of a
negligible amount of nuclear recoil and a very forward-emitted single pion.
Both CC $\pi^+$ and NC $\pi^0$ coherent production is possible. Many 
measurements of these cross sections exist, but mostly at higher 
energies~\cite{vilain}. Surprisingly, K2K observes no evidence 
for CC coherent pion production and sets an upper limit on this cross section.
This intriguing result is the first experimental measurement of 
neutrino-induced coherent pion production at low energy. \\

In addition to their new coherent pion results, K2K also presented 
an update of their axial mass ($M_A$) fits from QE data in their near
detectors~\cite{sakuda}. K2K's $M_A$ measurement is the first determination
of this parameter on a water target and is part of an important effort
aimed at improving our current knowledge of QE cross sections and
kinematics on nuclear targets. \\

Preliminary CC scattering results from the MiniBooNE experiment were also 
shown for the first time. Using a simple requirement that 
events contain two decay electrons in the final state, MiniBooNE isolates
a clean sample of CC $\pi^+$ interactions in mineral oil. This sample 
constitutes roughly five times more CC $\pi^+$ data than all previous data 
sets combined. Combining this sample with comparably high statistics 
QE data taken from the same experiment, MiniBooNE reports the first 
measurement of the ratio of  CC $\pi^+$/QE cross sections at low energy 
on a nuclear target~\cite{jocelyn}. The measured ratio is roughly consistent 
with current model predictions, given the large uncertainties in the
theory. In the future, MiniBooNE anticipates measurements of the NC $\pi^0$ 
and coherent pion production cross sections, which will complement those 
from K2K. \\

In addition to neutrino measurements, electron scattering data is also
a necessary input towards improving low energy neutrino simulations.
Because neutrino oscillation experiments commonly employ heavy targets to 
ensure high event yields, they rely on accurate nuclear model predictions. 
Higher statistics and more precise electron-nucleus scattering measurements 
are crucial for constraining and discriminating between such 
models (see for example~\cite{sakuda-summary}). Dedicated low $Q^2$ electron
data, recently taken by the JUPITER experiment, comprise part of a 
collaborative program linking the electron and neutrino physics communinities.
A very preliminary look at these data was provided at this workshop with 
the goal towards having $5\%$ cross section measurements for use by the 
neutrino community in the near future~\cite{bodek}.

\section{Future Initiatives}

Talks on future neutrino cross section opportunities were featured in the
second part of the session on experimental results. Antineutrino running at 
MiniBooNE was one such possibility~\cite{morgan}. Given the scarcity of 
antineutrino cross section measurements at low energy, such data will be 
especially important for future CP violation searches in the neutrino sector. 
In addition, a proposal for bringing the fine-grained K2K SciBar detector to 
the Fermilab Booster neutrino beamline was presented along with a discussion 
of its physics potential~\cite{morgan}. If approved, this would allow 
continued measurement of $\sim1$ GeV neutrino interactions making use of 
a pre-existing fine-grained detector and intense low energy neutrino 
source.  \\

Further, the recently approved MINERvA experiment will provide dedicated,
high statistics neutrino scattering physics measurements starting in 
2008~\cite{minerva}. The experiment features a fully active, 
fine-grained near detector placed in the newly commissioned NuMI beamline 
at Fermilab. This detector will uniquely allow measurements of neutrino 
interactions across a very broad range of energies (the NuMI beam can run 
in low, medium, and high energy modes). MINERvA can perform an impressive 
list of neutrino scattering studies with unprecedented detail, ranging from 
measurements of axial form factors, QE scattering, NC and CC resonance 
production, coherent pion production, strange particle production, deep 
inelastic scattering, and parton distributions~\cite{minerva}. The 
incorporation of multiple nuclear targets will additionally allow a mapping 
of nuclear effects in neutrino interactions for the first time as well as 
provide important information on the $A$ dependence of the coherent pion 
cross section. Given its broad potential for further improving our current 
knowledge, the construction and operation of this experiment is eagerly 
awaited. \\

The working group also expanded our discussion to include possible future
neutrino scattering measurements at very low energies ($\sim5-100$ MeV), the
region of interest for nuclear model builders and 
nuclear astrophysicists. Proposals for options to build a neutrino facility 
at the Spallation Neutron Source (SNS)~\cite{stancu} and to make use of low 
energy beta beams were discussed~\cite{volpe}. Such measurements would help 
span an important gap between between reactor and acclerator-based 
neutrino-nucleus measurements, where little data currently exists. \\

Finally, the potential for neutrino scattering measurements using extremely 
high resolution liquid argon detectors was presented~\cite{bonnie} along with 
some new ideas on how to relate such measurements to those on more 
conventional targets~\cite{lar}. The use of liquid argon could clearly 
allow more accurate determination of the various exclusive neutrino 
channels and increase the precision of neutrino cross section measurements 
beyond current capabilities.

\section{Questions for NuFact06}

Both the experimental and theoretical presentations in Working Group 2
focused primarily on surveying the current landscape, thereby providing
a sense of where we might be in the coming years. Future NuFact workshops 
must now focus on where to go from here, specifically addressing which 
additional neutrino scattering measurements are needed. Several questions
along these lines were presented at the workshop and are posed here:

\begin{itemize}
  \item After the complete set of anticipated results from the K2K, MiniBooNE,
        and MINERvA experiments have arrived, what additional neutrino 
        scattering measurements can and should be made?
  \item Is there specific information needed by future neutrino oscillation
        experiments that we will not already have in hand? Joint
        Working Group 1 and 2 sessions should focus on this issue.
  \item What other neutrino beams should be investigated? Consider
        the need for additional antineutrino measurements, 
        hadron production data, direct measurements of unoscillated $\nue$ 
        cross sections, and narrow band beams for dedicated NC studies.
  \item What more can be gained from employing additional neutrino targets? 
        This includes new, high statistics, light target measurements 
        (off hydrogen and deuterium) and the use of polarized targets.
  \item Are there other detector technology improvements that should
        be studied and pursued in addition to liquid argon detectors?
\end{itemize}

Now that we have a better understanding of where we are headed in
neutrino scattering physics in the short term, a concerted effort 
focused on additional measurements, beams, and detectors we might 
need beyond this is clearly necessary. 

\section{Conclusions}

New data from K2K and MiniBooNE already help fill the gap in our
present knowledge of low energy neutrino cross sections. This workshop saw
several new results from these experiments, including measurements 
in regions where there were previously no data. Future neutrino
oscillation experiments hoping for greater precision will require even
tighter control of these cross sections. In this 
regard, we look forward to further advancement that can be achieved at 
the MINERvA experiment and potentially from stopped pion sources, beta 
beams, and liquid argon detectors. While past workshops (including this one)
have commonly concentrated on the current state of affairs in neutrino 
scattering physics, we encourage future workshops to project their focus 
further into the future.

\section{Acknowledgements}

The author thanks J. Morfin and fellow co-convenor, M. Sakuda, with whom 
it was a pleasure to organize this session. 
I would also like to thank all of the Working Group 2 speakers and 
participants, and extend sincere thanks to the NuFact05 organizers for 
their very generous travel support, which made it possible for many of 
our working group speakers to attend this workshop.


\end{document}